\documentclass[aps,pra, twocolumn, amsmath, amssymb, showpacs, superscriptaddress]{revtex4}

\usepackage{graphicx}% Include figure files
\usepackage{bm}% bold math
\usepackage{xcolor}
\usepackage{ulem}

\begin{document}

\title{Universal methods for suppressing the light shift in atomic clocks using power modulation}

\author{V. I. Yudin}
\email{viyudin@mail.ru}
\affiliation{Novosibirsk State University, ul. Pirogova 1, Novosibirsk, 630090, Russia}
\affiliation{Institute of Laser Physics SB RAS, pr. Akademika Lavrent'eva 15B, Novosibirsk, 630090, Russia}
\affiliation{Novosibirsk State Technical University, pr. Karla Marksa 20, Novosibirsk, 630073, Russia}
\author{M.~Yu.~Basalaev}
\affiliation{Novosibirsk State University, ul. Pirogova 1, Novosibirsk, 630090, Russia}
\affiliation{Institute of Laser Physics SB RAS, pr. Akademika Lavrent'eva 15B, Novosibirsk, 630090, Russia}
\affiliation{Novosibirsk State Technical University, pr. Karla Marksa 20, Novosibirsk, 630073, Russia}
\author{A. V. Taichenachev}
\affiliation{Novosibirsk State University, ul. Pirogova 1, Novosibirsk, 630090, Russia}
\affiliation{Institute of Laser Physics SB RAS, pr. Akademika Lavrent'eva 15B, Novosibirsk, 630090, Russia}
\author{J.~W.~Pollock}
\affiliation{National Institute of Standards and Technology, Boulder, Colorado 80305, USA}
\affiliation{University of Colorado, Boulder, Colorado 80309-0440, USA}
\author{Z.~L.~Newman}
\affiliation{National Institute of Standards and Technology, Boulder, Colorado 80305, USA}
\author{M.~Shuker}
\affiliation{National Institute of Standards and Technology, Boulder, Colorado 80305, USA}
\author{A.~Hansen}
\affiliation{National Institute of Standards and Technology, Boulder, Colorado 80305, USA}
\author{M.~T.~Hummon}
\affiliation{National Institute of Standards and Technology, Boulder, Colorado 80305, USA}
\author{R.~Boudot}
\affiliation{FEMTO-ST, CNRS, UBFC, ENSMM, 26 rue de l'$\acute{e}$pitaphe 25030 Besancon, France}
\author{E.~A.~Donley}
\affiliation{National Institute of Standards and Technology, Boulder, Colorado 80305, USA}
\author{J.~Kitching}
\affiliation{National Institute of Standards and Technology, Boulder, Colorado 80305, USA}
%

%\date{\today}

\begin{abstract}
We show that the light shift in atomic clocks can be suppressed using time variation of the interrogation field intensity. By measuring the clock output at two intensity levels, error signals can be generated that simultaneously stabilize a local oscillator to an atomic transition and correct for the shift of this transition caused by the interrogating optical field. These methods are suitable for optical clocks using one- and two-photon transitions, as well as for microwave clocks based on coherent population trapping or direct interrogation. The proposed methods can be widely used both for high-precision scientific instruments and for a wide range of commercial clocks, including chip-scale atomic clocks.
\end{abstract}

%\pacs{}

\maketitle

\section{Introduction}
Atomic clocks have had broad impact in both fundamental physics and practical applications. They have enabled, for example, some of the most precise tests of general relativity \cite{Delva_2018,Herrmann_2018} and put stringent limits on the possible variation with time of fundamental constants \cite{Safronova_2019}. In addition, they underlie many aspects of modern technological infrastructure such as global navigation satellite systems (GNSS) and high-speed telecommunications systems.  The current state of the art, problems, and some prospects for the further development of atomic clocks of various types and purposes are well-presented in reviews \cite{Shah_2010,Poli_2013,Ludlow_2015}.

The most important metrological characteristics of an atomic clock are its long-term stability and accuracy. One of the main factors limiting these performance metrics is the ac Stark shift of atomic levels caused by the presence of any optical fields used to probe the atoms. The ac stark shift is particularly problematic for atomic clocks  based on coherent population trapping, and optical clocks, where optical fields are necessarily present in the interrogation sequence. Understanding, and ultimately suppressing, this light shift is therefore important in improving the long-term frequency stability of these types of references. Particular successes in this direction have been achieved over the past decade for atomic clocks based on Ramsey spectroscopy. This process was begun in Ref.~\cite{yudin2010}, where the method of so-called hyper-Ramsey spectroscopy was developed, which made it possible to reduce the light shift and its variations for a reference atomic resonance by several orders of magnitude \cite{hunt12,huntemann2016}. Further development of the hyper-Ramsey approach has used new phase variants to build an error signal \cite{NPL_2015,Zanon_2014,Zanon_2016,Yudin_2016,Zanon_2015,Zanon_2017,Zanon_2018}. This allows for significant improvement in the efficiency of suppression of the light shift in atomic clocks.

Auto-balanced Ramsey spectroscopy (ABR) is another effective approach that was first experimentally demonstrated in a $^{171}$Yb$^+$ ion clock \cite{Sanner_2017}. This approach was rigorously substantiated and generalized in Ref.~\cite{Yudin2018_PRAppl}, and also recently realized in atomic clocks based on coherent population trapping (CPT) \cite{Hafiz_2017,Shuker2019_PRL}. In ABR, two Ramsey sequences with different Ramsey times are used. A primary control loop stabilizes the clock frequency as in conventional Ramsey spectroscopy, while a second loop controls an adjustable property of the first and/or second Ramsey pulses, for example a phase jump of the local oscillator during the Ramsey sequence. An alternative method named combined error signal (CES) spectroscopy has been recently proposed \cite{Yudin2018_NJP} and demonstrated \cite{Shuker2019_APL}.  In contrast to the ABR protocol, the Ramsey-CES method uses a single combined error signal, constructed by subtracting the error signals obtained from the two Ramsey sub-sequences with an appropriate normalization factor (calibration coefficient). This one-loop method offers light-shift mitigation and reduces the complexity of implementing the two-loop control system required for ABR-like protocols at the cost of having to measure or otherwise establish the normalization factor.

Although successful as laboratory demonstrations, these new approaches have had no significant effect on the development and improvement of commercial atomic clocks, which typically use continuous wave (CW) spectroscopy rather than Ramsey spectroscopy. The main reason is that from the viewpoint of commercial instrumentation, reliability and relative simplicity of devices play a critical role and CW spectroscopy in atomic vapor cells is preferred. Moreover, in a number of important cases Ramsey spectroscopy is impractical because of the short lifetime of the excited state. For example, this is true for modern optical clocks using vapor cells with alkali atoms, such as for one-photon sub-Doppler spectroscopy with counterpropagating waves at the S$\leftrightarrow$P transition (e.g., see Ref.~\cite{Hafiz_2016}); two-photon spectroscopy on the S$\leftrightarrow$D transition (e.g., see Refs.~\cite{Perrella2013_PRA,Perrella2013_OL,Martin2018,Martin2019}); and Doppler-free spectroscopy of molecular iodine (e.g., see Ref.~\cite{Schuldt_2017}). In addition, in atomic clocks based on Ramsey spectroscopy in atomic beam and fountain (e.g., see in Ref.~\cite{Riehle_book}), the high-efficient ABR and CES methods developed in Refs.~\cite{Sanner_2017,Yudin2018_PRAppl,Yudin2018_NJP} also cannot be used, because in these devices the free evolution time cannot be varied to form two Ramsey sequences with different Ramsey times.

While CW spectroscopy is more widely used in commercial clocks at present, general methods to mitigate the light shift in a manner similar to the ABR and CES protocols has not yet been developed. To date, several approaches to the problem of the light shift in CW spectroscopy are known for various types of clocks. The first efforts in this direction were made for microwave CPT clocks, for which the light shift can be suppressed by a suitable choice of the rf modulation index of the laser field \cite{Zhu2000,Shah2006}. Light shifts have also been suppressed in conventional optically pumped microwave clocks by the appropriate choice of the frequency $\omega_\text{OP}$ of the optical pumping field \cite{Happer2009}. Recently, a method has been proposed that can suppress the light shift for two-photon spectroscopy on the transition $(5s\,$$^{2}S_{1/2})\leftrightarrow (5d\,$$^{2}D_{5/2})$ in $^{87}$Rb through the use of two interrogating laser fields at different frequencies \cite{Gerginov2018}. Therefore, the development of new effective and universal methods of suppressing the light shift for CW spectroscopy of clock transitions is currently a very relevant research topic.

In this paper, we develop two methods for suppressing the light shift and its fluctuation in atomic clocks based on either CW or Ramsey spectroscopy. Both methods use a power modulation (PM) which is sequential alternating operations with two different laser powers $P_1$ and $P_2$. The first method operates with a single combined error signal (PM-CES), constructed by subtracting the error signals obtained from the two sub-sequences ($P_1$ and $P_2$) with an appropriate normalization factor (calibration coefficient) dependent on the ratio $P_1/P_2$. The second method uses a two-loop approach to feed back on and stabilize the clock frequency $\omega$ as well as a second (concomitant) parameter $\xi$, which determines the value of the artificial ``anti-shift'' of the clock transition that actively auto-compensates the original shift (we refer to this as ACS). The operation of PM-ACS consists of the correlated stabilization of both variable parameters $\omega$ and $\xi$, which leads to the light shift cancellation for the clock frequency $\omega$. Another variant of PM-ACS using low frequency harmonic power modulation is also proposed. While the two-loop PM-ACS is more complicated to implement than the single-loop PM-CES, it also requires fewer constraints as outlines below. Both the PM-CES and the PM-ACS methods can be applied in optical clocks using one-photon and two-photon spectroscopy, as well as in rf clocks based on CPT resonances and optical pumping clocks.

\begin{figure}[t]
\centerline{\scalebox{0.75}{\includegraphics{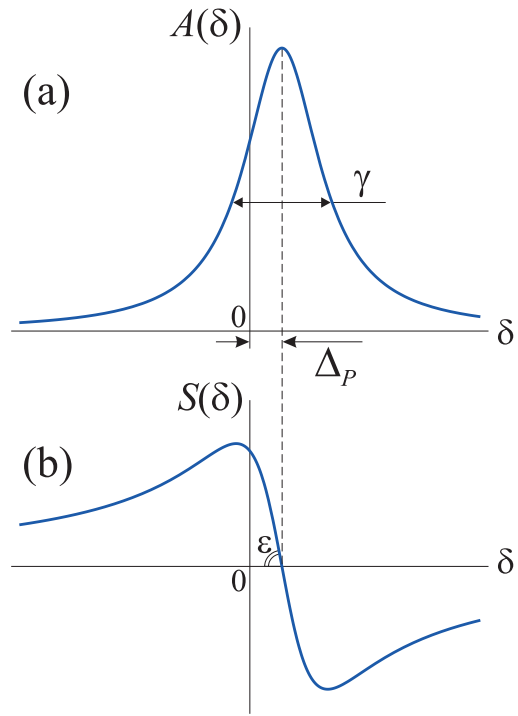}}}\caption{
Illustration of characteristic spectroscopic lineshapes. (a) the resonant lineshape of the spectroscopic signal $A(\delta)$ at the clock transition; (b) the corresponding error signal $S(\delta)$.} \label{Fig1}
\end{figure}

\section{General description of the light shift in atomic clocks}

Let us consider an atomic clock (either optical or microwave ranges) in which the frequency of the local oscillator $\omega$ is stabilized by a reference atomic transition with an unperturbed frequency $\omega_0$. For this purpose, a resonant spectroscopic signal with a characteristic linewidth $\gamma $ [see Fig.~\ref{Fig1}(a)] is used, on the basis of which the dispersion-like error signal $S(\delta)$ is generated in some known way (e.g., by the use of harmonic frequency modulation or frequency jumps) as a function of the detuning $\delta=\omega-\omega_0$ [see Fig.~\ref{Fig1}(b)]. The standard operation of an atomic clock is to use a feedback loop to stabilize the frequency $\omega$ at the zero of the error signal:
\begin{equation}\label{S_0}
S(\delta)=0\,,
\end{equation}
which, in the ideal case, corresponds to $\delta=0$, i.e. $\omega=\omega_0$.

However, under the influence of a laser field, the resonant atomic transition experiences a light shift $\Delta_P$ [see Fig.~\ref{Fig1}(a)], which also manifests in the error signal [see Fig.~\ref{Fig1}(b)]. In this case, the frequency stabilization can be mathematically represented as
\begin{equation}\label{S_shift}
S(\delta-\Delta_P)=0\,,
\end{equation}
which leads to the following result
\begin{equation}\label{omega_shift}
\delta=\Delta_P\;\Rightarrow\; \omega=\omega_0+\Delta_P\,,
\end{equation}
describing effects of the light shift on the stabilized frequency. We assume further that the main source of light shift is the ac Stark shift, which negatively affects not only the accuracy of atomic clocks, but also their long-term stability due to temporal variations in the value of $\Delta_P(t)$ caused by power and frequncy fluctuations of the laser field. Moreover, for rf clocks based on coherent population trapping, additional variations of $\Delta_P(t)$ can also occur from fluctuations in the modulation index of the laser field at the operating rf frequency $\omega$.

Below we consider two methods for suppressing the ac Stark shift and its variations in atomic clocks.

\section{Combined error signal at power modulation (PM-CES)}

In this section, we describe a method of the combined error signal (PM-CES). We will assume that the atomic clock operates in a time-interleaved manner at two different values of the probe laser field power, $P_1$ and $P_2$. For definiteness, we assume $P_2> P_1$. Moreover, we suppose that the following conditions are fulfilled:

1. The light shift is linear in the optical power:
\begin{equation}\label{shift_linear}
\Delta_P=cP\,,
\end{equation}
where $c$ is an empirical coefficient.

2. The slope of the error signal in the line center has a pure power-law dependence on $P$ [see Fig.~\ref{Fig1}(b)]:
\begin{equation}\label{err_slope}
\tan\varepsilon=bP^{\alpha}\,,
\end{equation}
where $b$ and $\alpha$ can be arbitrary. This assumption implies that the system cannot exhibit saturation or power broadening of the spectroscopic transition.

3. The light shift is much smaller than the linewidth:
\begin{equation}\label{shift_width}
\Delta_P\ll \gamma\,.
\end{equation}

\begin{figure}[t]
\centerline{\scalebox{0.5}{\includegraphics{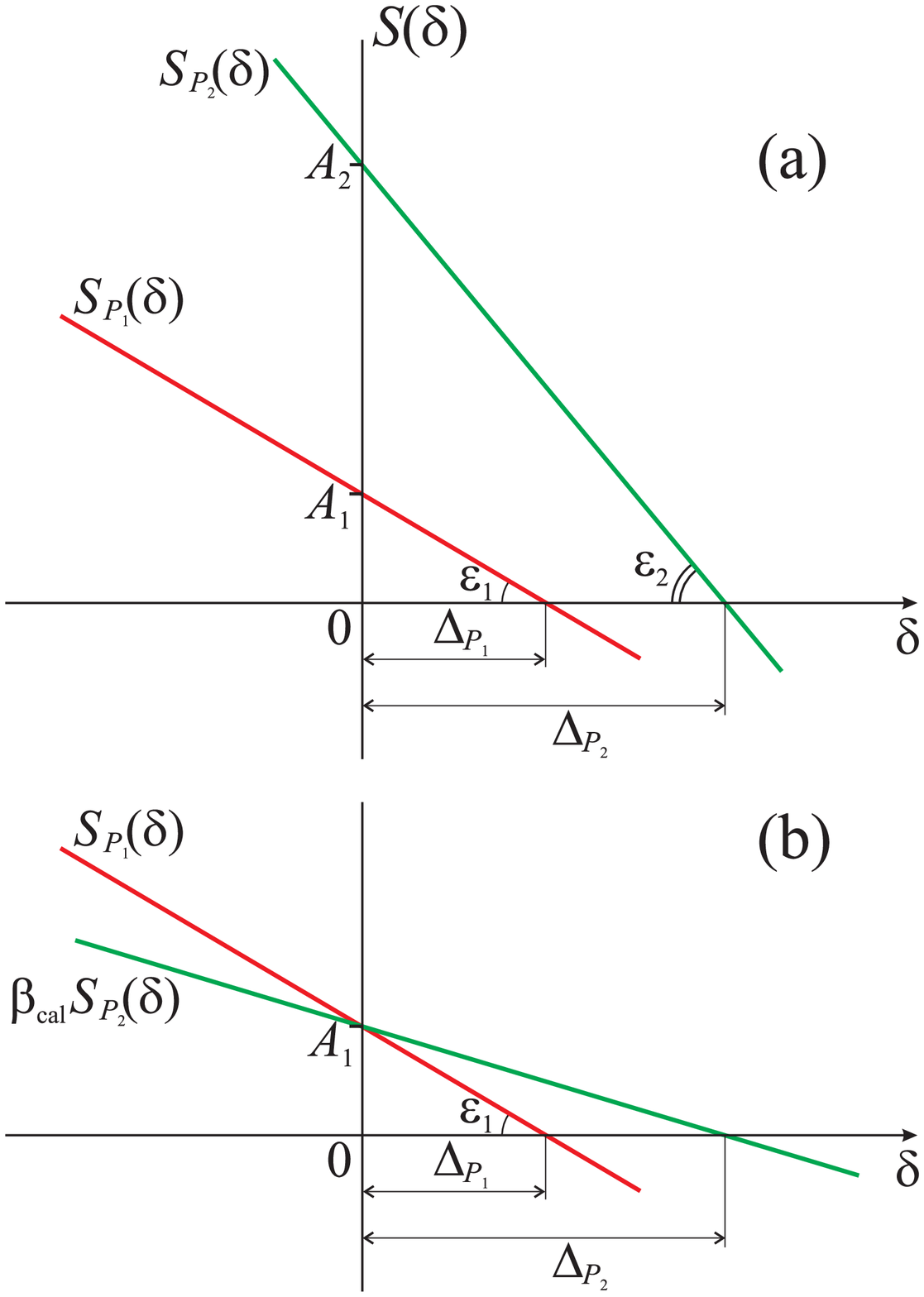}}}\caption{
Illustration of error signals. (a) linear central sections of the error signals $S_{P_1}(\delta)$ (red line) and $S_{P_2}(\delta)$ (green line) taking into account light shifts $\Delta_{P_1}$ and $\Delta_{P_2}$;
(b) transformation of the upper figure (a) when instead of the dependence $S_{P_2}(\delta)$, the product $\beta_\text{cal}S_{P_2}(\delta)$ is presented (green line), in which the calibration coefficient $\beta_\text{cal}$ [see Eq.~(\ref{beta})] is used.} \label{Fig2}
\end{figure}

\begin{figure}[t]
\centerline{\scalebox{0.5}{\includegraphics{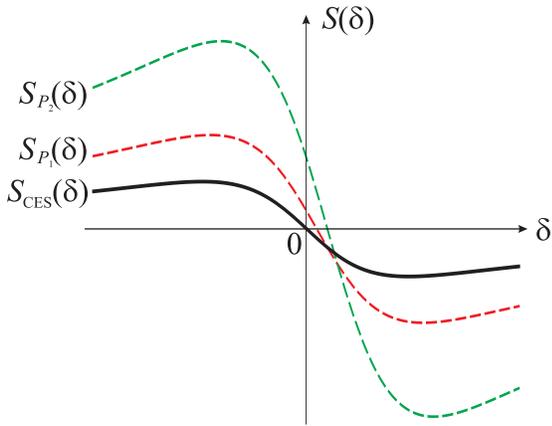}}}\caption{
Illustration showing the absence of a light shift for the combined error signal $S_\text{CES}(\delta)$ (black solid line), while the common error signals $S_{P_1}(\delta)$ (red dashed line) and $S_{P_2}$ (green dashed line) experience light shifts.} \label{Fig3}
\end{figure}

Let us consider now the error signals $S_{P_1}(\delta)$ and $S_{P_2}(\delta)$ for two values $P_1$ and $P_2$. Because of condition (\ref{shift_width}), near the zero of the error signals, the linear approximation can be used with good accuracy, as shown in Fig.~\ref{Fig2}(a). Using well-known trigonometric formulas, we find the ratio of the segments $0A_2$ and $0A_1$ in Fig.~\ref{Fig2}(a):
\begin{equation}\label{A2A1}
\frac{0A_2}{0A_1}=\frac{\Delta_{P_2}\tan\varepsilon_2}{\Delta_{P_1}\tan\varepsilon_1}=\left(\frac{P_2}{P_1}\right)^{\alpha+1},
\end{equation}
where we used expressions (\ref{shift_linear}) and (\ref{err_slope}). If we multiply the error signal $S_{P_2}(\delta)$ by the calibration coefficient $\beta_\text{cal}$, which is inverse to the ratio (\ref{A2A1}):
\begin{equation}\label{beta}
\beta_\text{cal}=\left(\frac{P_1}{P_2}\right)^{\alpha+1},
\end{equation}
then the dependence $\beta_\text{cal}S_{P_2}(\delta)$ will cross another error signal $S_{P_1}(\delta)$ at the point $A_1$ [as shown in Fig.~\ref{Fig2}(b)], which is located on the vertical axis corresponding to the unshifted frequency, $\delta=0$. As a result, if we construct a combined error signal (PM-CES) as
\begin{equation}\label{CES}
S_\text{CES}(\delta)=S_{P_1}(\delta)-\beta_\text{cal}S_{P_2}(\delta)\,,
\end{equation}
then the dependence $S_\text{CES}(\delta)$ will cross the horizontal axis at the point $\delta=0$, as shown in Fig.~\ref{Fig3}. Thus, if we use PM-CES (\ref{CES}) to stabilize the frequency $\omega$, as a result we get:
\begin{equation}\label{CES_0}
S_\text{CES}(\delta)=0\;\Rightarrow\;\delta=0\;\Rightarrow\;\omega=\omega_0\,,
\end{equation}
which means that the atomic clock light shift and its variations are suppressed. In addition, to maximize $S_\text{CES}(\delta)$, the condition $P_2\gg P_1$ should hold.

\begin{figure}[t]
\centerline{\scalebox{0.28}{\includegraphics{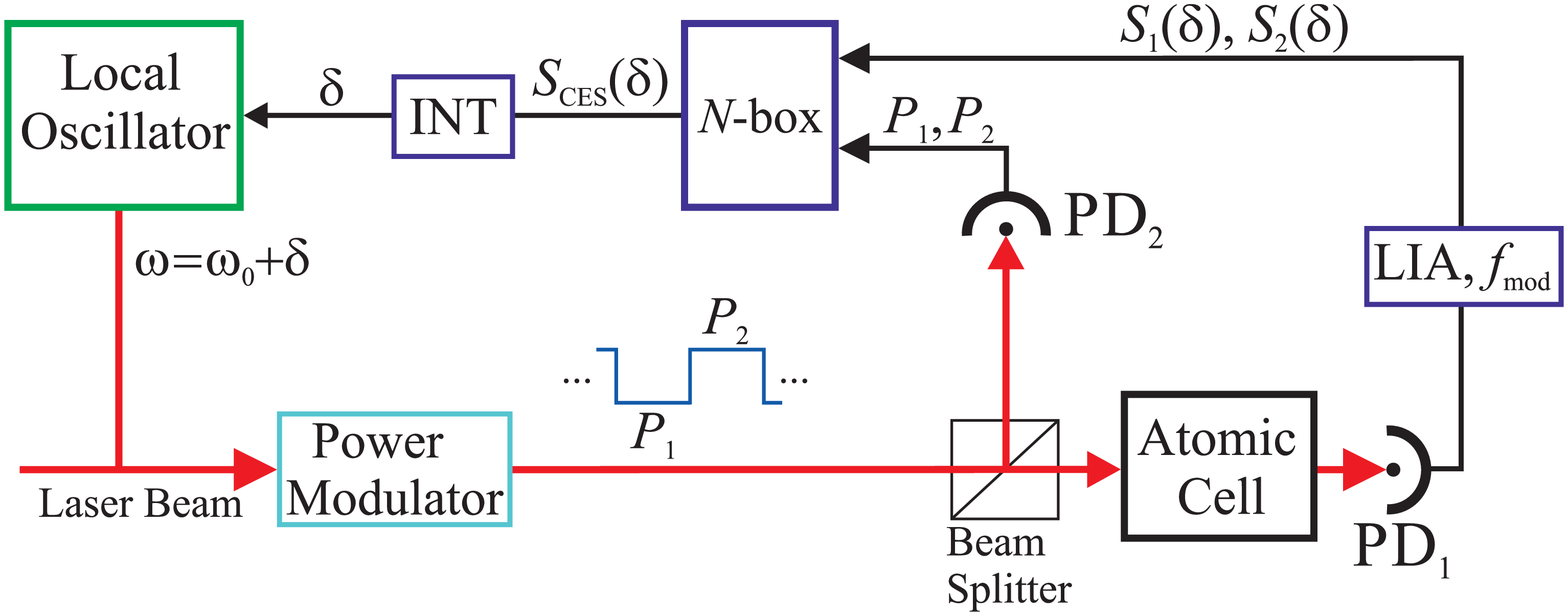}}}
\caption{The general scheme of the experimental implementation of the PM-CES method. INT, integrator; LIA, lock-in amplifier; PD, photodetector.} \label{Fig4}
\end{figure}

A potential experimental implementation of PM-CES is presented in Fig.~{\ref{Fig4}}. The power modulator creates an alternating sequence of two different powers $P_1$ and $P_2$ of the laser. The frequency of power modulation is much slower than $f_\text{mod}$, the modulation frequency used to generate the error signals $S_{P_{1,2}}$ (see Fig.~{\ref{Fig4}}, lock-in-amplifier). Note that, for example, an acousto-optical modulator or a liquid crystal waveplate followed by a polarizer can be used as the power modulator. Next, before entering the atomic cell, a beam-splitter is installed, which splits the laser beam into two. One beam passes through the atomic cell and forms a spectroscopic signal and error signals $S_{P_1}(\delta)$ and $S_{P_2}(\delta)$ using the first photodetector PD$_1$. In the case of one-photon spectroscopy, this can be a transmission signal, while in the case of two-photon spectroscopy, this can be a fluorescence signal. The second beam is incident on photodetector PD$_2$ and is used to calculate the ratio $P_1/P_2$, i.e., to determine the calibration coefficient $\beta_\text{cal}$ [see Eq.~(\ref{beta})]. The digital block ``$N$-box'' based on data from the photodetectors PD$_1$ and PD$_2$ generates the combined error signal $S_\text{CES}(\delta)$ [see Eq.~(\ref{CES})], which is then used to stabilize the frequency $\omega$ of the local oscillator (LO) by the condition $S_\text{CES}(\delta)=0$.

PM-CES is a quite general method suitable for various types of spectroscopy. For example, for one-photon spectroscopy in counterpropagating waves and in the weak saturation regime of the clock transition, the signal size is linear in power and  we have $\alpha=1$. In this case, the calibration coefficient is:
\begin{equation}\label{1_photon}
\beta_\text{cal}^\text{(1-ph)}=\left(\frac{P_1}{P_2}\right)^{2}.
\end{equation}
In the case of a fluorescence signal for two-photon spectroscopy, where the signal size scales with the power squared, $\alpha=2$, the calibration coefficient is:
\begin{equation}\label{2_photon}
\beta_\text{cal}^\text{(2-ph)}=\left(\frac{P_1}{P_2}\right)^{3}.
\end{equation}
PM-CES can also be used in CPT clocks as well as for optical pumping clocks. In general, the success of the PM-CES method is directly related to how accurately the basic conditions in Eqs.~(\ref{shift_linear})-(\ref{shift_width}) are fulfilled in experiments.

In addition, in some cases, to obtain information about the laser power $P$, we can also use the first photodetector PD$_1$. In this case, the presence of a second photodetector PD$_2$ is optional, which provides some technical simplification. Note also that the PM-CES method presented above can be considered as a PM analog of the Ramsey-CES method developed in Ref.~\cite{Yudin2018_NJP}.

\section{Auto-compensation of the light shift (PM-ACS)}
In this section, we develop a method to auto-compensate the light shift (ACS) in spectroscopy using a power modulation (PM-ACS). Unlike the PM-CES method developed in the previous section, PM-ACS only requires the linearity of the light shift Eq.~(\ref{shift_linear}), while the other two conditions Eqs.~(\ref{err_slope}) and (\ref{shift_width}) do not matter. In PM-ACS, we assume the use of an additional frequency shifter, which for any laser power $P$ allows us to shift the frequency of the local oscillator $\omega$ by the value $\xi P$, where $\xi$ is a well-controlled and variable parameter. This shift $\Delta_\text{ACS}=\xi P$ plays the role of an artificial anti-shift, which allows us to completely compensate for the actual light shift $\Delta_P=cP$ and its fluctuations. Indeed, taking into account the actual shift $\Delta_P$ and the artificial anti-shift $\Delta_\text{ACS}$, the result of the frequency stabilization with the use of the error signal can be represented as a solution of the equation
\begin{equation}\label{Serr_anty}
S_P(\delta,\xi)=S_P(\delta+\xi P-cP)=0
\end{equation}
for an unknown $\delta$. Then, for $\xi=c$ we get $\delta=0$, i.e. $\omega=\omega_0$.

Below we present the implementation of the PM-ACS method in two different modifications: for stepwise and harmonic modulations of the laser power.

\subsection*{a) PM-ACS for stepwise modulation of power}
We will assume that the atomic clock operates at two different values of the probe laser field power $P_1$ and $P_2$. The PM-ACS method consists of the following repeating cycles. For interrogation with the laser power $P_1$, the parameter $\xi$ is fixed, and we stabilize the variable detuning $\delta$ (i.e., LO frequency $\omega$) at the zero point of the error signal: $S_{P_1}(\delta,\xi_\text{fixed})=0$. The operation of this servo loop can be presented as the following recurrent sequence:
\begin{equation}\label{servo_omega}
\delta_{n}=\delta_{n-1}+r\cdot S_{P_1}(\delta_{n-1},\xi_\text{fixed})\,,
\end{equation}
where $r$ is a feedback factor for the frequency loop, and $n$ is the step index counter for the sequence. After this procedure, we switch to interrogation with power $P_2$, where we fix the previously-obtained detuning $\delta$ and stabilize the variable parameter $\xi$ at the zero point of the second error signal: $S_{P_2}(\delta_\text{fixed},\xi)=0$. The operation of this second servo loop can be presented as another recurrent sequence:
\begin{equation}\label{servo_xi}
\xi_{m}=\xi_{m-1}+q\cdot S_{P_2}(\delta_\text{fixed},\xi_{m-1})\,,
\end{equation}
where $q$ is a feedback factor for $\xi$-loop, and $m$ the step index counter for the second servo loop.

If we continue these cycles, then the final result consists of the stabilization of both parameters, $\delta=\bar{\delta}_\text{clock}$ and $\xi=\bar{\xi}$, which corresponds to the solution of a system of two equations:
\begin{eqnarray}\label{syst1}
&& S_{P_1}(\delta,\xi)=S_{P_1}(\delta+\xi P_1-cP_1)=0\,, \nonumber \\
&& S_{P_2}(\delta,\xi)=S_{P_2}(\delta+\xi P_2-cP_2)=0\,,
\end{eqnarray}
for two unknowns $\delta$ and $\xi$. The value $\bar{\delta}_\text{clock}$ describes the frequency shift of the atomic clock. Assuming $S_P(x)\propto  x$ for $x \approx 0$, the Eqs.~(\ref{syst1}) are equivalent to the system given by
\begin{equation}\label{syst2}
\delta+(\xi-c)P_1=0\,,\quad \delta+(\xi-c)P_2=0\,,
\end{equation}
which has the solution
\begin{equation}\label{solution}
\delta=0\,,\quad \xi=c\,.
\end{equation}
Thus, we have shown that the PM-ACS method always leads to zero shift of the stabilized frequency $\omega$ in an atomic clock, $\bar{\delta}_\text{clock}=0$.

\begin{figure}[t]
\centerline{\scalebox{0.27}{\includegraphics{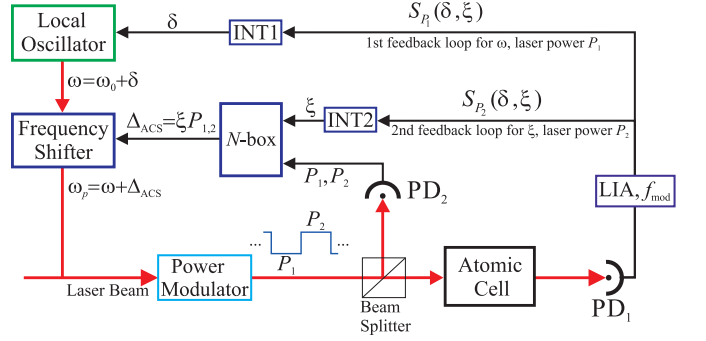}}}
\caption{The general scheme of the experimental implementation of the CW-ACS method for stepwise modulation of power. INT, integrator; LIA, lock-in amplifier; PD, photodetector.} \label{Fig5}
\end{figure}

Fig.~\ref{Fig5} shows a schematic diagram of the implementation of the PM-ACS method, which has two feedback loops: a main loop to stabilize the local oscillator detuning $\delta$, and an auxiliary loop to stabilize the scale factor $\xi$ for the artificial anti-shift  $\Delta_\text{ACS}$.  In the absence of the auxiliary loop ($\xi = 0$), the main loop locks the local oscillator to the light shifted atomic resonance, $\omega_0+cP_{1,2}$. When the auxiliary loop is enabled, feedback from the error signal $S_{P_2}(\delta,\xi)$ drives the scale factor $\xi$ to be non-zero, and correspondingly causes the probe frequency $\omega_p=\omega + \xi P_2$ to be locked to the light shifted resonance, $\omega_0+cP_2$.  This condition, when applied to both operating powers, $P_1$ and $P_2$, is equivalent to Eqs.~(\ref{syst2})-(\ref{solution}).

Note that the continuous use of two feedback loops (for $P_1$ and $P_2$) will lead to some decrease of the short-term stability, because it increases the length of each cycle of frequency stabilization. This can be remedied in the following way. In the initial period of frequency stabilization, we use the two-loop PM-ACS method. This step allows us to determine the value of the concomitant parameter $\bar{\xi}$ with satisfactory accuracy. Then the procedure of long-term frequency stabilization can be done by only one feedback loop for the error signal $S_{P_1}(\delta,\bar{\xi})$ using the previously-determined parameter $\bar{\xi}$. Moreover, we can regularly (but rarely) use the two-loop PM-ACS again to remeasure $\bar{\xi}$ and correct any long-term drift. On the one hand, this approach allows for the regular adjustment of the parameter $\bar{\xi}$ (to eliminate, for example, an influence of possible slow variations of the characteristics of the second photodetector PD$_2$). On the other hand, because the measurement is intermittent, it will not lead to significant reduction of the signal-to-noise ratio and corresponding degradation of the long-term frequency stability. The choice of the ratio of $P_1$ to $P_2$ is an additional control that can be used to optimize the process.

PM-ACS is a general method useful for a variety of clocks laser spectroscopy, including optical clocks using one-photon and two-photon spectroscopy and microwave clocks based on optical pumping or CPT. We stress that the amount the light shift can be suppressed using the PM-ACS method is directly related to how accurately the linearity condition in Eq.~(\ref{shift_linear}) is met. In this context, it is very relevant to continue a detailed study of light shifts for various atomic clocks in order to determine the most suitable operating modes in which the linearity in Eq.~(\ref{shift_linear}) is best fulfilled.

We also note that the PM-ACS method can be considered conceptually as a PM analog of one of the variants of generalized auto-balance Ramsey spectroscopy Ref.~\cite{Yudin2018_PRAppl,Shuker2019_PRL}, in which an additional frequency shift was used as the concomitant parameter.

\subsection*{b) PM-ACS for low frequency harmonic modulation of power}
\begin{figure}[t]
\centerline{\scalebox{0.27}{\includegraphics{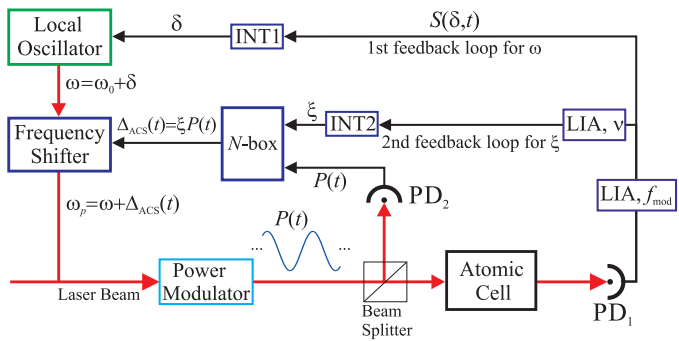}}}
\caption{The general scheme of the experimental implementation of the PM-ACS method for low-frequency harmonic modulation of laser power (\ref{mod}). INT, integrator; LIA, lock-in amplifier; PD, photodetector.} \label{Fig6}
\end{figure}

Fig.~\ref{Fig6} shows a schematic diagram of the implementation of another version of PM-ACS using low-frequency harmonic modulation of the laser power:
\begin{equation}\label{mod}
P(t)=P_0\left[1+M\sin (\nu t)\right],
\end{equation}
where $M<1$. In this case, both the actual shift $\Delta_P(t)=cP(t)$ and the artificial anti-shift $\Delta_\text{ACS}(t)=\xi P(t)$ become time-dependent according to a harmonic law with a frequency $\nu$. Note that the power modulation frequency $\nu$ is much lower than the frequency $f_\text{mod}$ of the frequency deviation used to generate the error signal $S(\delta)$, as in Ref.~\cite{Shah2006} for a CPT clock. Additionally we require that the bandwidth of the feedback to the local oscillator be much lower than the power modulation frequency $\nu$.

The operation of the scheme in Fig.~\ref{Fig6} is the following. The effect of power modulation in Eq.~(\ref{mod}) is monitored as modulation on the error signal $S(\delta,t)$ at the frequency $\nu$. This low-frequency modulation is observed on the error signal because a change in the total optical power produces a corresponding change in the atomic resonance frequency due to both the actual light shift $\Delta_P(t)$ and the artificial anti-shift $\Delta_\text{ACS}(t)$:
\begin{equation}\label{shifts12}
\Delta_P(t)-\Delta_\text{ACS}(t)=(c-\xi)P(t)\,.
\end{equation}
Then, the condition $\xi=c$ at which the power shift contribution in Eq.~(\ref{shifts12}) becomes zero occurs when the $\nu$-frequency modulation on the error signal vanishes.

\section{Conclusion}
We have described two different methods for suppressing the light shift and its variation in atomic clocks based on either CW or Ramsey spectroscopy. The first method, PM-CES, uses only one feedback loop to stabilize the LO frequency but requires the measurement of a calibration factor to create a suitable error signal. The second method (PM-ACS) method requires two loops but has fewer restrictions on the size of the light sift and nature of the error signal.These methods are general in that they can be applied to optical clocks using one-photon and two-photon spectroscopy, as well as microwave clocks based on CPT resonances and optical pumping. The implementation of the PM-CES and PM-ACS techniques can in principle lead to significant improvement of the accuracy and long-term stability for high-precision and commercial atomic clocks.

In addition, although the schemes depicted in Figs.~\ref{Fig4} and \ref{Fig5} are more related to the CW spectroscopy, we believe that the corresponding schemes for Ramsey spectroscopy (in an atomic beam or fountain) will not cause difficulties.

\begin{acknowledgments}
We thank Sihong Gu, Leo Hollberg, and Joseph Christesen for useful discussions and Yun-Jhih Chen and Abijith Kowligy for helpful comments on the manuscript.
This work was supported by the Russian Scientific Foundation (no. 16-12-10147). V. I. Yudin and A. V. Taichenachev were also supported by the Russian Foundation for Basic Research (no.~20-02-00505 and no.~18-02-00822). This work is also partially funded by NIST.
\end{acknowledgments}

\end{document}